\documentclass[aps,pra,superscriptaddress,twocolumn,groupedaddress,showpacs,amsmath,amsfonts,floatfix]{revtex4}
\usepackage{graphicx}
\usepackage{bm}
\usepackage{color}

\begin{document}
\title{Controlled turbulence regime of electron cyclotron resonance ion source  \\ for improved multicharged ion performance}

\author{V.\,A.\,Skalyga}
\author{I.\,V.\,Izotov}
\author{A.\,G.\,Shalashov}
\author{E.\,D.\,Gospodchikov}
\author{E.\,M.\,Kiseleva}
\affiliation{Institute of Applied Physics of the Russian Academy of Sciences, 603950, 46 Ulyanova st., Nizhny Novgorod, Russia}
\affiliation{Lobachevsky State University of Nizhny Novgorod, 603950, 23 Gagarina ave., Nizhny Novgorod, Russia}
\author{O.\,Tarvainen}
\affiliation{STFC ISIS Neutron and Muon Source, Rutherford Appleton Laboratory, Didcot OX11 0QX, United Kingdom}
\author{H.\,Koivisto}
\author{V.\,Toivanen}
\affiliation{Department of Physics, University of Jyv\"askyl\"a, FI-40014 Jyv\"askyl\"a, Finland}

\date{\today}

\begin{abstract} 
Fundamental studies of excitation and non-linear evolution of kinetic instabilities of strongly nonequlibrium hot plasmas confined in open magnetic traps suggest new opportunities for fine-tuning  of conventional electron cyclotron resonance (ECR) ion sources. These devices are widely used for the production of particle beams of high charge state ions. Operating the ion source in controlled turbulence regime allows increasing the absorbed power density and therefore the volumetric plasma energy content in the dense part of the discharge surrounded by the ECR surface, which leads to enhanced beam currents of high charge state ions. We report experiments at the ECR ion source at the JYFL accelerator laboratory, in which adopting of a new approach allows to increase the  multicharged ion beam current up to two times, e.g.\ to 95 $\mu$A of $\mathrm{O}^{7+}$ achieved with mere 280 W power at 11.56 GHz. A theoretical model supporting and explaining the experimental findings is presented. 
The study suggests that the controlled turbulence regime has the potential to enhance the beam currents of modern high-performance ion sources, including state-of-the-art superconducting devices.
\end{abstract}


\maketitle
\section{\label{sec:intro}Introduction}

Ion sources with plasma heating under the electron cyclotron resonance (ECR) condition are used in nuclear physics research as injectors of multicharged ions into accelerators. The majority of remarkable results in this area over the past several decades derives from significant progress of the ECR ion sources. The main trend in such development is increasing the frequency and power of the microwave radiation used for plasma heating that allows operation with denser plasma at higher power loads. However, a  number of more subtle techniques have been successfully applied to boost specific features, such as increasing the current and average charge of the beam or increasing the current of high charge state ions, without making critical changes to the construction of the system. Those methods include gas mixing \cite{b1}, two-frequency heating \cite{b2}, fine-tuning of the microwave frequency \cite{b3}, wall coatings \cite{o1}  as well as preglow \cite{b4} and afterglow \cite{b5} effects in a pulsed mode of operation. In this paper, we report on a novel approach that allows to increase the high charge state oxygen beam currents up to two times with the possibility to achieve even better gain with heavier elements.
This approach resulted from fundamental plasma physics studies of Poincar\'e--Andronov--Hopf bifurcations in coherent (maser) emission from strongly heated nonequilibrium plasma confined in a mirror trap \cite{prl}.
%

A traditional approach for designing ECR ion sources is based on using semi-empirical scaling laws proposed by Geller \cite{b6}. The main one dictates that the ion current corresponding to the maximum of the ion charge state distribution is proportional to the square of the heating radiation frequency or, equivalently, of the magnetic field strength in the ECR zone: 
$I_{\mathrm{peak}} \propto \omega_{\mathrm{ECR}}^2 \propto B_{\mathrm{ECR}}^2 $.
In order to achieve ion confinement times on the order of 1--10 ms and to suppress magnetohydrodynamic instabilities, the magnetic field of a classical ECR source uses a minimum-$B$ configuration, which is a combination of the solenoid field of a simple mirror trap and a multipole (usually sextupole) radial field. Minimum-$B$ configuration can be characterized by the maximum  field on the axis at the microwave injection end $B_{\mathrm{inj}}$, the field on the wall of the plasma chamber $B_{\mathrm{rad}}$, the maximum field at the beam extraction aperture $B_{\mathrm{ext}}$, and the minimum field in the trap center $B_{\mathrm{min}}$. Generally considered, there is an  optimal combination of these four parameters for the ECR source performance that scales with the resonant field as \cite{b7}
\begin{equation}\label{2nd}
\frac{B_{\mathrm{inj}}}{B_{\mathrm{ECR}}} \approx 4, \;
\frac{B_{\mathrm{rad}}}{B_{\mathrm{ECR}}} \approx 2, \;
\frac{B_{\mathrm{ext}}}{B_{\mathrm{ECR}}} \approx 1.8, \;
\frac{B_{\mathrm{min}}}{B_{\mathrm{ECR}}} \approx 0.8.
\end{equation}
It has been established experimentally that when the magnetic field is increasing up to a level determined by \eqref{2nd}, there is a steady growth in the average charge of the extracted ions \cite{b8}. On the contrary, when the field values exceed the values determined by \eqref{2nd}, the average charge begins to decrease with increasing field, and the ion beam current starts to oscillate periodically.

The negative effect of increasing the magnetic field  was explained by the appearance of periodic, $10^0-10^3$ Hz, development of electron cyclotron instabilities of the plasma that excited when ${B_{\mathrm{min}}}/{B_{\mathrm{ECR}}}$ is above the optimal value defined by \eqref{2nd} \cite{b9,b10}. Due to the nature of ECR heating, a strongly non-equilibrium electron distribution is formed in such plasma, possessing a significant fraction of fast electrons with relativistic energies (up to 1 MeV), in addition to the ``cold'' and ``warm'' electron populations \cite{o3}.  Under certain conditions, such distribution becomes unstable, which causes periodic bursts of coherent microwave plasma emission accompanied with abrupt ejections of energetic electrons from the magnetic trap. The latter in turn provoke significant fluctuation of the plasma potential (up to values of more than  +1 kV \cite{b11} from the steady state value of 10--50 V.) and generate powerful pulsed X-ray radiation when the electrons collide with the chamber wall. This process can be explained by the fact that resonant fast electrons may efficiently transfer their transverse (to the magnetic field) kinetic energy to the unstable electromagnetic waves and, thus, scatter into the loss cone leaving the trap. Such periodic eruptive  dynamics are well described within the paradigm of a cyclotron maser formed in a highly non-equilibrium ECR discharge plasma \cite{b12maser}. Due to the large jumps of the plasma potential, the electrostatic confinement of multicharged ions in the potential well formed in the trap center (like in a classical ambipolar trap \cite{b13}) is interrupted, which leads to a decrease in the cumulative ion confinement time and subsequently to a strong modulation of the ion current with a period of the plasma instability. Thus, the range of strong magnetic fields, in particular $B_{\mathrm{min}}$, that significantly exceed the values determined by condition \eqref{2nd}, has always been considered unfavourable for the generation of multicharged ions.

\begin{figure}[tb]
\includegraphics[width=83mm]{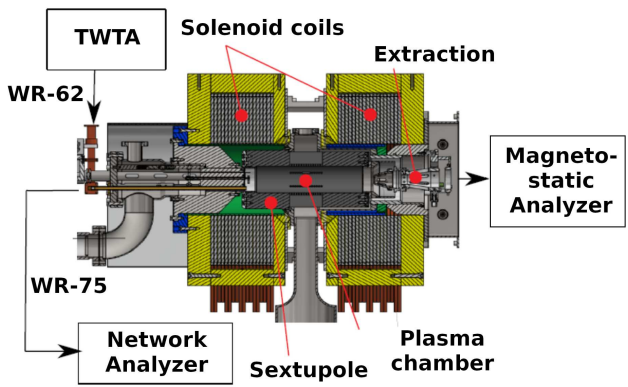}
\caption{ \label{fig1} The experimental setup.
}\end{figure}

During  studies of fundamental processes of plasma-wave interactions performed with a 10--14 GHz A-ECR-U type ion source at the Accelerator Lab of the University of Jyv\"askyl\"a (JYFL) \cite{b15}, it was demonstrated that the pulse-periodic mode of the plasma cyclotron maser can be controllably converted to a continuous-wave (cw) mode characterized by relatively weak and steady microwave emission and flux of energetic electrons precipitating from the trap ends \cite{prl}. Such cw regime is observed in a narrow range of strong axial magnetic fields above the threshold of the transition from stable plasma to the pulse-periodic kinetic instability regime. Since high-power microwave bursts and electron showers do not occur in the cw mode, oscillations of the plasma potential are damped, and therefore, the confinement of long-living highly charged ions is not affected. In this paper, we study the generation of multicharged ions in this particular, previously unexplored, operation mode of the ion source above the traditional upper limit  of the solenoid field strength \eqref{2nd}.

\section{Experimental results}

The experiments were carried out with oxygen plasma on the same setup at JYFL \cite{prl}. The schematic of the setup is shown in Fig.~\ref{fig1}. 
We used a travelling wave tube as source of the heating microwave radiation instead of the standard klystron source. This allowed operation at lower ECR frequency (tunable in the range of 10.8--12.4 GHz), which enabled using a wider range of magnetic fields relative to  $B_{\mathrm{ECR}}$ and, hence,  to cover a wider range of values in \eqref{2nd}. 
%

\begin{figure*}
  \includegraphics[width=130mm]{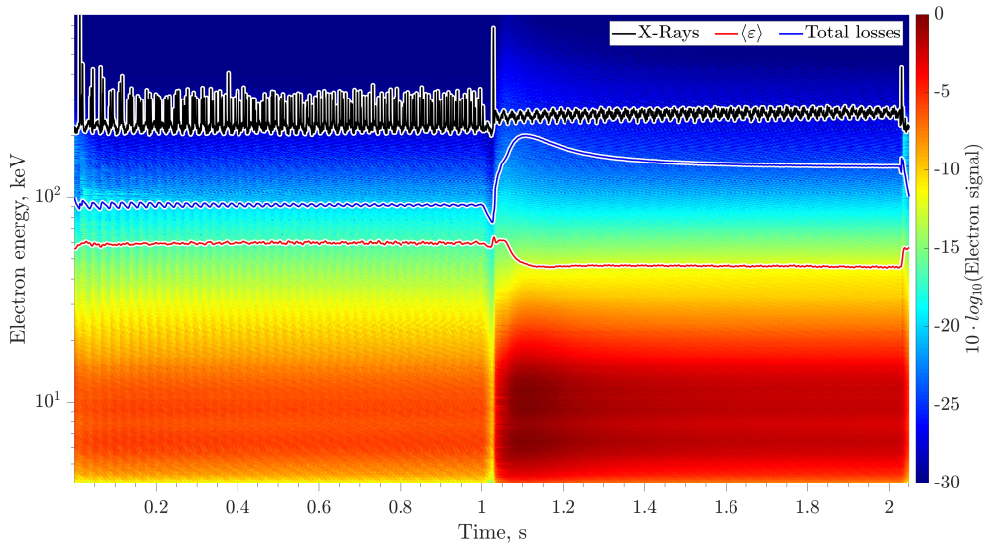}
  \caption{\label{fig2a} 
Transition from the pulse-periodic to cw emission mode while the ECR power increases from 15 to 50 watts: the signal of the X-ray detector (black line), the average energy of energetic electrons escaping the trap through the extraction (red line),  the energy distribution of the electrons leaving the trap (color map). Experiment is performed with the heating frequency of 12.0~GHz,  oxygen pressure $3.5\times10^{-7}$~mbar, and $B_{\mathrm{min}} / B_{\mathrm{ECR}}=0.982$. 
}
\end{figure*}

\begin{figure}
  \includegraphics[width=83mm]{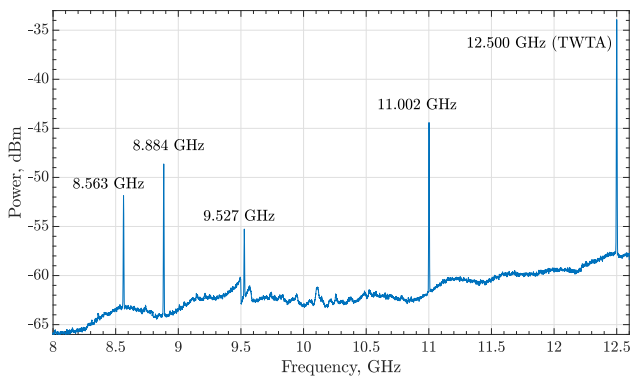}
  \caption{\label{fig2b}  
	Plasma emission spectrum in the cw  mode. Experiment is performed with the heating  power of 250~W and frequency of 12.5~GHz,  oxygen pressure $3.5\times10^{-7}$~mbar, and $B_{\mathrm{min}} / B_{\mathrm{ECR}}=0.945$. 
}
\end{figure}

The ion beam charge state distribution was determined with a magnetostatic spectrometer resolving the mass-to-charge values of the extracted ions. 
A novel technique of measuring the electron energy distribution escaping through the extraction aperture was used to study the correlation between fast electron losses and the ion source performance \cite{b16}. 
In contrast to \cite{prl}, where different modes of the cyclotron maser were studied with a broadband oscilloscope, here we used a more accessible scheme based on a spectrum analyzer operating in frequency scanning mode in the range from 8 to 12.5 GHz. When a discharge was forced into the cw mode, several stable lines of low-power radiation appeared in the plasma microwave emission spectrum. Additionally, an X-ray detector in the accumulation mode was used as a diagnostic of the X-ray power flux. The presence of a distinct pulse-periodic signal from the X-ray detector unambiguously indicated the pulse-periodic regime of the cyclotron instability.
Typical signals in our main diagnostic channels during the most interesting event, the transition from the pulse-periodic to cw mode, are illustrated in Fig. \ref{fig2a}. An example of a plasma emission spectrum in the cw mode is shown in Fig. \ref{fig2b}.

\begin{figure}
  \includegraphics[width=83mm]{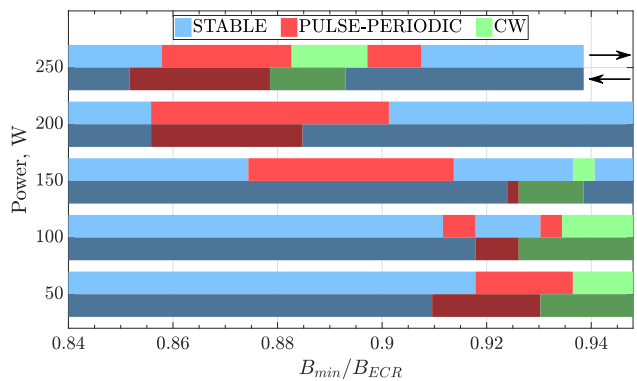}
  \caption{\label{fig2}
A diagram showing different regimes of a cyclotron maser in the magnetic-field-strength/ECR-power plane: stable plasma (blue),  pulse-periodic generation (red), cw generation (green). Shades indicate the direction of the slow variation of the magnetic field strength: increase (light bars) or decrease (dark bars). Experiment is performed with the heating frequency of 12.0 GHz, and oxygen pressure $3.5\times10^{-7}$~mbar. 
}
\end{figure}

Our setup allows controlled variation of the solenoid coil current during a continuous operation of the ion source. Thus, we used a traditional approach, 
that is scanning over a range of confining magnetic field strength while keeping all other controllable parameters, including heating frequency, heating power and neutral gas pressure, constant. 
At each measurement point we assured that plasma is settled to a stationary, although eventually bursty, state; no transient dynamics were recorded. This was straightforward since the setup allowed non-stop operation during the whole experimental campaign. Hence, the plasma discharge was continuously sustained for approximately two weeks, which allowed to avoid all problems with the conditioning and achieve nearly 100\%-reproducibility of the discharge regimes.

In terms of conditions \eqref{2nd}, we start with the optimal configuration and then vary proportionally $B_{\mathrm{inj}}$, $B_{\mathrm{ext}}$, and $B_{\mathrm{min}}$ while keeping  $B_{\mathrm{ECR}}$ and, approximately, $B_{\mathrm{rad}}$ constant; hereafter each magnetic configuration is characterized by a single parameter $B_{\mathrm{min}}/B_{\mathrm{ECR}}$. 
Slow variation of the magnetic field strength triggered a series of bifurcations in cyclotron instability dynamics, e.g., when the field strength was increased a stable plasma rapidly switched to the pulse-periodic regime, and finally to the cw regime. 
Figure \ref{fig2} shows sequences of such transitions measured at different levels of the microwave heating power measured in increasing or decreasing magnetic field. It is seen that the cw generation occurs in narrow islands (colored green). 
Note also a  hysteresis feature: the boundaries between the regimes depend on whether the magnetic field is increased or decreased. This subtle effect was predicted theoretically in \cite{epl,bible}, and is now observed experimentally for the first time. The regimes of cw emission were detected at almost any frequency of plasma heating that could be provided by our source. 
Note that the threshold value of $B_{\mathrm{min}} / B_{\mathrm{ECR}}$ above which the plasma is unstable, increases with decreasing heating power, as described in \cite{b17}. In experiments reported below, due to the relatively low heating power the threshold value was 0.85--0.9, which exceeds the classical optimal value predicted by \eqref{2nd}.  
The dependence of other transitions on the heating power, including the bifurcation feature, is quite complicated. Although a theoretical backgroung is developed \cite{epl,bible}, its verification is still a matter of future research.

\begin{figure*}[tb]
\includegraphics[width=173mm]{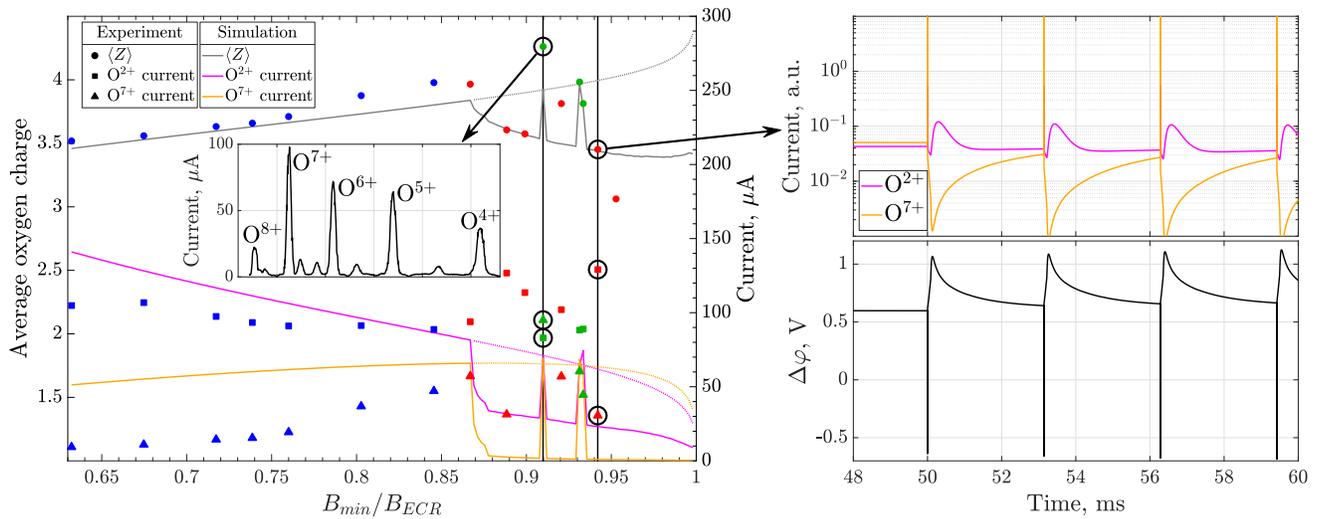}
\caption{ \label{fig3} Left plot: the dependence of  O$^{2+}$, O$^{7+}$ currents, and the value of the average charge of the extracted ion beam on the magnetic filed strength. The experimental data are shown by dots, different colors correspond to the stable plasma (blue), pulse-periodic (red), and cw (green) regimes of  the instability. The heating frequency is 11.56 GHz, the power is 280 W, the parameters of the magnetic system comply with the standard JYFL conditions.
Lines represent the results of modeling with (solid lines) and without (dotted lines) taking into account the instability. 
Inset: the measured ion beam spectrum corresponding to  $B_{\mathrm{min}} / B_{\mathrm{ECR}}=0.91$. Right plots: the calculated time-traces of O$^{2+}$,  O$^{7+}$ currents and the ambipolar potential dip $\Delta\varphi$, i.e.\ the difference between  potentials at the trap center  and in the magnetic plug, 
during the pulse-periodic instability with a repetition rate of 315 Hz which corresponds to $B_{\mathrm{min}} / B_{\mathrm{ECR}} = 0.945$.
}\end{figure*}


To study the prospects of implementing the cw regime for achieving increased currents of multicharged ions, 
the ion beam charge state distributions were measured in a varying magnetic field for the most commonly used settings (heating frequency, gas pressure); then the average charge state was calculated from the measured beam currents as a function of $B_{\mathrm{min}} / B_{\mathrm{ECR}}$. A representative example 
is shown in Fig.~\ref{fig3}. The obtained trends generally represent the well-known results described in \cite{b8}---the average charge and the current of high charge state ions (O$^{7+}$, but not O$^{2+}$) increase with the magnetic field up to certain threshold value of $B_{\mathrm{min}} / B_{\mathrm{ECR}}\approx 0.85$, corresponding to the transition from stable plasma to the pulse-periodic regime of the cyclotron instability.   In the pulse-periodic regime the high charge state production efficiency decreases with increasing $B_{\mathrm{min}} / B_{\mathrm{ECR}}$ until a sharp drop is encountered at $B_{\mathrm{min}} / B_{\mathrm{ECR}}\approx 0.95$. 

However, the trend changes completely when the system switches to the cw emission mode, see blue points in the range of 0.9--0.95. Here the values of the currents and average charge  return to the previous trend and even grow relative to the values at the threshold of transition to the unstable regime. In particular,  the O$^{7+}$ ion current doubles relative to the stable mode at $B_{\mathrm{min}} / B_{\mathrm{ECR}}=0.91$. 
It is important to note that the experimental points corresponding to the cw operation fit well the general trend of the growth of highly charged ion current with the increase of the magnetic field in the range corresponding to the stable plasma. 
This indicates that, unlike the pulse-periodic mode, in the cw regime the instability does not affect the ion confinement.
The full ion beam spectrum obtained in the best performance shot with the cw mode is shown in inset. It can be seen that the O$^{7+}$ current is 95 $\mu$A for the heating frequency of 11.56 GHz and a power of 280 W, which is a record yield per the unit of power deposited into the plasma for this ion source. 

\section{Theoretical model}

To verify our interpretation of the effect, the experimental conditions were simulated with a dedicated numerical model \cite{b19} successfully  applied and confirmed in many previous experiments. This model, based on a system of ionization and power balance equations, is developed to describe the dynamic processes of step-by-step electron impact ionization and accumulation of multicharged ions in the ECR discharge plasma confined in a mirror magnetic trap with taking into account the complex form of the electron distribution function and the dependence of particle lifetimes on the self-consistent ambipolar electric potential, which dominates the confinement of highly charged ions. This model has proven itself demonstrating a quantitative agreement with experimental data from various ion sources, such as the Phoenix family\cite{b20}, the high-current source SMIS-37 \cite{b21}, and, in particular, the JYFL A-ECR-U source used in the present study \cite{b22}. Good agreement between the model and experiment was demonstrated not only for continuous operation of the sources, but also in dynamic pulsed modes, describing the preglow and afterglow effects \cite{b23}. In this communication, we use (first time) the model to describe the effect of instabilities.  


As noted above, the decrease in the average charge of ions in the beam in the pulse-periodic instability mode is associated with the destruction of ion confinement due to jumps in the ambipolar potential stimulated by the pulsed ejection of hot electrons from the trap, absent in the cw mode. To describe this, we impose periodic degradation of the electron confinement caused by their interaction with the unstable waves and consequent ejection into the loss cone. Then the model resolves, iteratively at each time step, the quasi-neutrality condition for the plasma losses, $N_{e}/\tau_{e}=\sum Z_i N_i/\tau_i$, adjusting the ion lifetimes $\tau_i\propto\exp(eZ_i\Delta\varphi/T_i)$ by varying of the ambipolar potential dip $\Delta\varphi$. In this way, rapid electron losses result in switching from the  confinement of multicharged ions at $\Delta\varphi>0$ to the regime of its intense losses  which grow exponentially with the ion charge at $\Delta\varphi<0$. 
Therefore, the more effectively a high charge state ion is confined electrostatically in the stable plasma, the faster it escapes the trap during the instability burst.  

However, the electron confinement time $\tau_{e}$ is calculated self-consistently, so we are not able to define the function $\tau_{e}(t)$ explicitly, instead we impose modulation of the external technical parameter of the model---the {maximal} value of electron lifetime $\tau_{\mathrm{max}}(t)$.  
Usually this level is set to a large value of the order of 1 ms which is rarely reached in simulations.
But to simulate the rf-driven electron losses when the instability develops, we lower the upper boundary of the electron confinement time according to the following model:  
$\tau_{\mathrm{max}}=\max(\tau_\mathrm{min},\alpha\tau_\mathrm{stable}/E(t))$, where 
$\tau_\mathrm{min}$ is the {minimal} possible lifetime value, $\tau_\mathrm{stable}$ is the electron lifetime calculated without taking into account the instability, and $E(t)$ is a  dimensionless function proportional to the intensity of a periodically excited wave turbulence, 
\begin{equation}\label{4th}
E(t)=\frac{(D / T) \sqrt{1+(D / T)^{2}}}{\cos ^{2}(\pi t / T)+(D / T)^{2}},\;\;\frac1T\int_{0}^{T} E(t) \:\mathrm{d} t\equiv 1.
\end{equation}
Here we assume that the rf-driven electron losses are proportional to the wave intensity and the instability consists of a sequence of approximately Lorentz pulses with characteristic pulse length $D$ and repetition period $T$. The form of $E(t)$ is chosen to ensure the unit average value for all $T$ and $D$, this property is required for consistency with the cyclotron maser model \cite{epl,bible}.  For large enough $E(t)$ we have $\tau_{\mathrm{max}}=\tau_\mathrm{min}$, thus we impose the \emph{maximal} electron losses. The latter are defined by plasma removal with the ion-acoustic velocity $\upsilon_\mathrm{is}$ resulting in the gas-dynamic scaling $\tau_{\mathrm{min}}\approx L R/2\upsilon_\mathrm{is}$, where $L$ is the length between the mirrors, and $R$ is the mirror ratio.
Numerical coefficient  $\alpha\approx2$ is adjusted from the assumption that, for a sufficiently large period of quiet plasma between the instability spikes,  calculated currents should tend to the values in the stable plasma.  

An example of numerical simulation is shown in Fig.~\ref{fig3} together with the experimental data. The characteristic  duration of a single instability pulse was set to $D\approx1\;\mu$s according to our previous measurements \cite{b17}, the  pulse repetition period was taken from the experimental measurements and varied in the range of $T=3-30$ ms for the pulsed-periodic mode and set $T\sim D$ for the cw mode. 
Calculations clearly reproduce the negative effect of the pulse-periodic instability on the ion beam composition. At the same time,  under stable conditions, the average charge increases with an increase in the magnetic field due to enhancement of the ion confinement and the heating efficiency (with a decreasing volume of the ECR zone). At singular points corresponding to the cw instability, the ion beam parameters are close to those in the stable plasma.

The right panels in  Fig. \ref{fig3} show the simulated evolution of the ion currents and the ambipolar potential dip during the pulse-periodic instability. 
At the beginning of the instability development,  O$^{7+}$ current increases sharply, since ion lifetime rapidly decreases in a self-consistent manner following the electron lifetime, while O$^{2+}$ current is not affected so strongly. 
The sharp peak of high charge state ions has been verified experimentally \cite{b11}.
%
A short burst is then followed by the relaxation of the ion currents to the quasi-stationary state. The discharge evolves starting from the consecutive ionization of the neutrals and low-charged ions followed by a slow increase in the highly-charged ion content. A broad peak of O$^{2+}$ current is a consequence of strong, by two orders of magnitude, peak in the density of neutrals caused by the ion recombination on the chamber walls during the initial stage of the instability. 

\section{Discussion}

\begin{table*}
\centering
\caption{Conditions of high charge state oxygen beam production with the JYFL A-ECR-U ion source in the high-$B$ continuous-wave emission mode and other high-performance ECR ion sources: plasma chamber length $L_{\textrm{p}}$ and diameter $D_{\textrm{p}}$, extraction aperture diameter $D_{\textrm{ext}}$, microwave heating frequency $f$ and power $P$, maser instability control parameter ${B_{\textrm{min}}}/{B_{\textrm{ECR}}}$, and the best beam currents $I$ for O$^{6+}$ and O$^{7+}$ ions.}
\label{comparison_table1}
\begin{tabular}{lc*{8}c}
\hline
\hline
Ion source & $L_{\textrm{p}}$ [mm]\;\; & $D_{\textrm{p}}$ [mm]\;\; & $D_{\textrm{ext}}$ [mm]\;\; & $f$ [GHz]\;\; & $P$ [kW]\;\; & ${B_{\textrm{min}}}/{B_{\textrm{ECR}}}$ \;\;& $I(\textrm{O}^{6+})$ [$\mu$A]\;\;& $ I(\textrm{O}^{7+})$ [$\mu$A] \;\;& Ref.\\
\hline
JYFL (cw) & 270 & 76 & 8 & 11.56 & 0.28 & 0.945 & 68 & 95 & - \\
VENUS & 520&141&10 & 28+18 & 8.0+1.9 & 0.56 & 4760 &1360 & \cite{Lyneis} \\
SECRAL-II & 420&126&12 & 28+18 & 4.5+0.5 & 0.65 & 6700 &1750 & \cite{Sun} \\
SuSI & 440&101&10 & 24 & 5.2 & 0.61 & 2200&800 & \cite{Machicoane} \\
HIISI & 400&105&8 & 18+14 & 2.0+0.25 & 0.66 & 1300 & 600 & \cite{Koivisto} \\
\hline
\hline
\end{tabular}
\end{table*}

\begin{table*}
\centering
\caption{Comparison of derived key parameters of high charge state oxygen beam production with the  JYFL A-ECR-U ion source (cw mode) to high performance ECR ion sources: total plasma volume $V_{\textrm{p}}$, volume enclosed by the ECR surface $V_{\textrm{ECR}}$, corresponding power densities $\rho_{\textrm{p}}$ and $\rho_{\textrm{ECR}}$, and volumetric efficiencies $\eta=I/\rho_{\textrm{ECR}}$ for O$^{6+}$ and O$^{7+}$ ion production. The volumetric efficiencies are calculated using the volume enclosed by the ECR zone to represent the plasma volume relevant for high charge state production.}
\label{comparison_table2}
\begin{tabular}{lc*{5}c}
\hline
\hline
Ion source & $V_{\textrm{p}}$ [cm$^{3}$]\;\;& $V_{\textrm{ECR}}$ [cm$^{3}$]\;\; & $\rho_{\textrm{p}}$ [W/cm$^{3}$] \;\;&  $\rho_{\textrm{ECR}}$ [W/cm$^{3}$] \;\;& $\eta(\textrm{O}^{6+})$ [$\mu$A/W/cm$^{3}$]\;\; & $\eta(\textrm{O}^{7+})$ [$\mu$A/W/cm$^{3}$]\\
\hline
JYFL (cw) & 1220 & 29 & 0.23 & 9.7 & 8.4$\cdot$10$^{-3}$ & 11.7$\cdot$10$^{-3}$  \\
VENUS & 8120 & 620 & 1.2 & 16.0 & 0.78$\cdot$10$^{-3}$ & 0.22$\cdot$10$^{-3}$ \\
SECRAL-II & 5240 & 470 & 0.95 & 10.6 & 2.9$\cdot$10$^{-3}$ & 0.74$\cdot$10$^{-3}$ \\
SuSI & 3500 & 390 & 1.5 & 13.3 & 1.1$\cdot$10$^{-3}$ & 0.39$\cdot$10$^{-3}$ \\
HIISI & 3140&360 & 0.72 &6.3 & 1.6$\cdot$10$^{-3}$ & 0.74$\cdot$10$^{-3}$ \\
\hline
\hline
\end{tabular}
\end{table*}

The presented results clearly demonstrate the possibility of a significant increase in the yield of highly charged ions from the ECR ion source by tuning it to the continuous generation (cw) mode of a kinetic electron cyclotron instability. The search for such regime can be cumbersome, and its characteristics may depend on the chosen approach path to the desired regime. 
For example, we failed to demonstrate the effect at a higher heating frequency of 12.7 GHz. At this frequency, there was no fundamental increase in the current of highly charged ions  with the tuning to the cw mode compared to the level reached in the stable plasma (although the ion production efficiency increased compared to the dip caused by the pulse-periodic instability regime). Therefore, a fine tuning of the heating frequency may be necessary for the most effective operation; such feature of the ECR ion sources is well known \cite{b3}.  
Moreover, the system can possibly appear to be very sensitive to the accuracy of tuning parameters, resulting that the long-term operation of the source in this mode can sometimes be difficult to implement due to the slow drift of parameters, associated, for example, with the gas recycling regimes shifts on the plasma chamber walls or heating of certain elements. Nevertheless, such fine tuning can be of key importance in the limiting operation modes of accelerators with the aim of demonstrating record-breaking parameters.

In order to place our results in appropriate context, we have compared the performance of the JYFL A-ECR-U ion source operating in the high-$B$ cw emission mode to record oxygen beam currents of state-of-the-art large volume ECR ion sources. The ion sources, their plasma chamber dimensions, extraction aperture diameters and operational parameters to achieve the quoted high charge state beam currents are listed in Table~\ref{comparison_table1}. Such comparison is not straightforward because the extraction aperture diameters of the sources range from 8 to 12~mm and the transport efficiencies of the adjacent beamlines could be different. For high charge state ions, e.g.\ O$^{6+}$ and O$^{7+}$, which originate from the plasma volume near the source axis~\cite{Panitzsch}, the effect of the extraction aperture size is considered small. The plasma chamber length is often ill-defined and we have therefore used the length between the axial magnetic field maxima as the chamber length for VENUS and SuSI. The key parameters derived from these values, namely the plasma volume enclosed by the cyclotron resonance zone for cold electrons, the power densities calculated from the microwave power and the chamber dimensions as well as the estimated volume enclosed by the ECR zone, and the volumetric efficiencies of high charge state oxygen ion beam production are shown in Table~\ref{comparison_table2}. For the JYFL A-ECR-U, the volume enclosed by the ECR zone was calculated numerically using accurate magnetic field simulations. For the other sources, the plasma volume was estimated by considering the axial and radial extremes of the resonance zone derived from the magnetic field profiles of each ion source. The resonance zone was then approximated by a 3D ellipsoid surface limiting the dense plasma volume, described, e.g., by Mascali et al.~\cite{Mascali}, which slightly underestimates the real volume and, thus, inflates the volumetric efficiency values (in comparison to the JYFL A-ECR-U). Some of the record beam currents have been achieved in the two frequency heating mode to stabilise the plasma. In these cases we have used the higher frequency ECR zone to define $B_{\textrm{min}}/B_{\textrm{ECR}}$ and the dense plasmoid boundary. Note that the plasma volumes in the table are just typical values for high charge state tune that naturally have some variation up to $\pm30\%$ depending on particular experimental conditions.

One can see that high-performance ion sources operating at higher frequency and microwave power, and thus at higher plasma density and volume, naturally provide high charge state currents exceeding those reported in the present communication by up to two orders of magnitude. However, operating the JYFL A-ECR-U ion source in the cw emission mode allows reaching the same power densities as typically required in the high power, large-volume, ion sources. In this sense, our results are relevant to plasma conditions of the superconducting ion sources. At the same time, the volumetric efficiency of the high charge state ions is superior in the JYFL A-ECR-U. 

Although it still needs experimental verification, our theoretical interpretation and modeling suggest that the stabilization of the kinetic instability would increase high charge state ion confinement in the 3rd and 4th generation ECR ion sources as well. It is worth noting that for the large-volume sources listed in Tables~\ref{comparison_table1} and \ref{comparison_table2}, the high charge state oxygen ion beam currents were not observed to saturate as a function of the microwave power. This means that the performance of the modern superconducting ion sources could probably be increased by operating them in the controlled turbulence regime with high $B_{\textrm{min}}$ and high heating power density. Since the negative effect of burst-like instability acts on the beam currents through the ambipolar electrostatic potential, its influence increases exponentially with the charge state. Therefore, our technique might work even better for heavy ions such as xenon, which are more relevant for actual applications of high-performance sources, than for the oxygen beams presented here. 

It is worth mentioning that smaller, so-called 2nd generation ECR ion sources, similar or even identical to the source described in this work, are widely used in accelerator laboratories around the world, such as ANL, LBNL, KVI, INFN, JINR. In many cases, the increase of high charge state ion current even by 10\% is highly desirable, for example when beams are produced from expensive isotopes for long periods of time, e.g.\ for superheavy element research, allowing to either minimize the material consumption or reduce the experimental time. Then, our approach may be applied as operation parameters optimization (in a range not considered before) suitable for existing systems. This would potentially allow improving performance of many similar devices without extra expenses for building a new system.

\section{Conclusion}
We find that  operation of the ECR ion source in the cw mode allows to increase $B_{\textrm{min}}$ above $0.8B_{\textrm{ECR}}$, which is known to be the semi-empirical limit for the confining magnetic field strength optimized for high charge state production \cite{b8}, while maintaining plasma stability. Our result suggests that the scaling law described in Ref.~\cite{b8} is set by the threshold for the cyclotron maser instability. At $B_{\textrm{min}}/B_{\textrm{ECR}}>0.8$, the controlled turbulence regime offers a constant energy loss mechanism in the form of microwave emission and hot electron precipitation without abrupt destruction of the ion confinement, and leads to high plasma energy density and sufficiently long ion confinement time enabling high charge state production.

As a final remark we note that the reported phenomena of bifurcations of cyclotron instability dynamics have much in common with similar processes in space cyclotron masers realized in planet magnetospheres and other astrophysical objects \cite{trh}. Unlike native systems, the laboratory experiment with a compact open trap allows controllable tuning of the unstable plasma conditions by adjusting the magnetic configuration or/and a frequency and power of radiation used for plasma heating and, therefore, is well suited for studying subtle effects, like ambiguous dependence on the bifurcations on the control parameters reported in this paper, that are very hard to investigate in space  masers. 


\begin{acknowledgments}
We thank Dr. Janilee Benitez, Dr. Liangting Sun and Dr. Guillaume Machicoane for providing data and pointing us to correct references for the oxygen beam currents and ion source operational parameters displayed in Tables~\ref{comparison_table1} and \ref{comparison_table2}. 
This work was supported by the Russian Science Foundation (project No.\ 19-12-00377); authors also acknowledge the Academy of Finland for supporting the experimental campaign (project  No.\ 315855). 
\end{acknowledgments}

\end{document}